\documentclass[%
reprint,
superscriptaddress,
showpacs,
preprintnumbers,
nofootinbib,
amsmath,
amssymb,
amsfonts,
aps,
prd,
floatfix,
showkeys
]{revtex4-1}

\RequirePackage{atbegshi}

\usepackage[utf8]{inputenc}
\usepackage[T1]{fontenc}
\usepackage[ngerman,english]{babel}
\usepackage[matha]{mathabx} % To use better looking arrows
\usepackage[dvipsnames, svgnames, table]{xcolor}
\usepackage[braket, qm]{qcircuit}
\usepackage{bbm}
\usepackage{%
    graphicx,
    siunitx,
    multirow,
    booktabs,
    bbold,    % for matrix one
    pifont,   % for checkmarks
    capt-of,
    pgfplotstable,
    todonotes,
}
\usepackage{subfigure}     % to have subfigures
\renewcommand{\thesubfigure}{(\alph{subfigure})}
\makeatletter
  \renewcommand{\@thesubfigure}{\thesubfigure\space}
  \def\@currentlabel{\p@subfigure\thesubfigure}
\makeatother
\usepackage{hyperref}
\usepackage{braket}
\hypersetup{
    colorlinks=true,
    citecolor=black,
    linkcolor=black,
    urlcolor=black,
    anchorcolor=black,
    linktocpage
}
\usepackage{outlines}
\usepackage{etoolbox} % for \appto
\usepackage{cleveref} % To be loaded last!
\makeatletter%   https://tex.stackexchange.com/a/266344 needs etoolbox
\appto{\appendix}{%
  \@ifstar{\def\theequation@prefix{A.}}%
          {}%
}
\makeatother
\crefname{figure}{Figure}{Figures}
\crefname{table}{Table}{Tables}
\crefname{equation}{Eq.}{Eqs.}
\crefname{section}{Section}{Sections}
\crefformat{appendix}{the #2Appendix#1#3} % For single Appendix use this to avoid space after label
\graphicspath{{./figures/}}

\DeclareMathOperator{\Tr}{Tr}

\newcommand{\Nf}{N_\text{f}}
\newcommand{\clqcd}{CL\kern-.25em\textsuperscript{2}QCD}
\newcommand{\beq} {\begin{eqnarray}}
\newcommand{\eeq} {\end{eqnarray}}
\newcommand{\nn}{ \nonumber}
\newcommand{\eref}[1]{Eq.~(\ref{#1})}

\newcommand{\calH}{\mathcal{H}}
\newcommand{\calN}{\mathcal{N}}
\newcommand{\calZ}{\mathcal{Z}}
\newcommand{\bareT}{aT}
\newcommand{\bareMu}{a\mu_B}
\newcommand{\bareI}{a\mu_I}

\def \Lat {{\Lambda}}
\def \LatB {{\Lambda_B}}
\def \LatM {{\Lambda_M}}

\def \LatSpatB {{\Sigma_B}}
\def \LatSpatM {{\Sigma_M}}
\def \sumLatSpat {{\sum_{\{\LatSpatM,\LatSpatB\}}}}

\def \U {\rm U}
\def \SU {\rm SU}

\newcommand{\meson}{\mathfrak{m}}
\newcommand{\baryon}{\mathfrak{b}}
\newcommand{\hadron}{\mathfrak{h}}
\newcommand{\spin}{\mathfrak{s}}
\newcommand{\Hil}{{\mathbbm{H}_\hadron}}
\newcommand{\Id}{{\mathbbm{1}}}

\newcommand{\Nt}{N_\tau}
\newcommand{\at}{a_\tau}
\def \nn {\nonumber\\}
\newcommand{\lr}[1]{\left(#1\right)}
\def \hmu {\hat{\mu}}
\newcommand{\GC}{\delta_{\sum\limits_\mu k_\mu(x),3}}
\newcommand{\Exp}[1]{\exp \lr{#1}}

\begin{document}

\title{
Quantum Gate Sets for Lattice QCD in the strong coupling limit: $N_f=1$
}

\author{Michael Fromm}
 \email{mfromm@itp.uni-frankfurt.de}
 \affiliation{
  Institut f\"{u}r Theoretische Physik, Goethe-Universit\"{a}t Frankfurt\\
 Max-von-Laue-Str.\ 1, 60438 Frankfurt am Main, Germany
}

\author{Owe Philipsen}
 \email{philipsen@itp.uni-frankfurt.de}
 \affiliation{
  Institut f\"{u}r Theoretische Physik, Goethe-Universit\"{a}t Frankfurt\\
 Max-von-Laue-Str.\ 1, 60438 Frankfurt am Main, Germany
}

\author{Wolfgang Unger}
 \email{wunger@physik.uni-bielefeld.de}
 \affiliation{
Fakult\"{a}t f\"{u}r Physik, Bielefeld University\\
33615 Bielefeld, Germany
}

\author{Christopher Winterowd}
\email{winterowd@itp.uni-frankfurt.de}
\affiliation{
 Institut f\"{u}r Theoretische Physik, Goethe-Universit\"{a}t Frankfurt\\
 Max-von-Laue-Str.\ 1, 60438 Frankfurt am Main, Germany
}

\begin{abstract}
We derive the primitive quantum gate sets to simulate lattice quantum chromodynamics (LQCD) in the strong-coupling limit with one flavor of massless staggered quarks. This theory is of interest for studies at non-zero density as the sign problem can be overcome using Monte Carlo methods. In this work, we use it as 
a testing ground for quantum simulations. The key point is that no truncation of the bosonic Hilbert space is necessary as the theory is formulated in terms of color-singlet degrees of freedom (``baryons'' and ``mesons''). The baryons become static in the limit of continuous time 
and decouple, whereas the dynamics of the mesonic theory involves two qubits per lattice site. Lending dynamics also to the ``baryons'' simply requires to use the derived gate set in its controlled version.
\end{abstract}

%\preprint{}

% \pacs{12.38.Gc, 05.70.Fh, 11.15.Ha}
\maketitle

\section{Introduction}

Recently, there has been considerable interest generated by proposals for quantum simulations of gauge theories in high-energy physics. This interest has spread to lattice gauge theory (LGT), and in particular, to lattice quantum chromodynamics (LQCD). This development promises to allow direct access to real-time physics as well as to the physics of dense nuclear matter from LQCD. Considerable progress has been made in the universal gate-based approach both in the methods needed \cite{Lamm2019,Klco2019,Ciavarella2021, Davoudi2020} as well as their applications to theories with both discrete and continuous gauge groups to varying levels of approximation.

Although from current estimates it seems that near-term, large-scale simulations of LQCD on quantum computers are infeasible \cite{Kan2021}, one can still define physically interesting problems which can be studied in the noisy intermediate-scale quantum (NISQ) era \cite{Preskill2018quantumcomputingin}. In the gate-based approach, one is of course forced to work in the Hamiltonian formulation. In the case of LGT, the Hamiltonian formulation appeared quite early \cite{Kogut1975} but was of secondary interest due to the success of Markov chain Monte Carlo methods which rely on the Lagrangian formulation. Another pioneering study identified the mapping of operators appearing in the Hamiltonian into spin operators for continuous gauge groups assuming a cutoff in the infinite-dimensional Hilbert space of the gauge bosons \cite{Byrnes2006}. This line of analysis has also been extended to fermions \cite{Zohar2015}. In order to alleviate the systematic effects of Hilbert space truncation, one can attempt to improve the Kogut-Susskind Hamiltonian by adding terms which reproduce the desired physics for a given cutoff. This has very recently been done for the strong-coupling limit using a similarity renormalization group approach \cite{Ciavarella2023}.  

The strong-coupling limit of lattice QCD has been a long-standing field of interest, both in the Wilson and staggered formulations.  Of particular interest were the reformulation of the theory in color-singlet degrees of freedom and their first numerical simulations~\cite{Rossi1984, Karsch1989}. These were mainly motivated by the possibility to either cure or 
alleviate the sign problem by the
reformulation, and hence study LGT systems at non-zero baryon density. Equipped with new algorithms, these studies were revisited some years later, leading to a complete mapping of the phase diagram for one staggered flavor in the strong coupling limit $(\beta = 0)$~\cite{Forcrand2010} and beyond~\cite{Forcrand2014}. All of the above-mentioned studies start from the Lagrangian formulation and thus, in principle, have little to say about real-time evolution of quantum states. Recently, a continuous-time partition function has been formulated leading to the construction of a Hamiltonian describing the physics of the strong-coupling limit with $N_f=1$ flavors of staggered fermions~\cite{Klegrewe2020a}. Further efforts were undertaken to extend this to the case of $N_f=2$~\cite{Pattanaik2021, Pattanaik2022}.

There exists then ample motivation to study the strong-coupling limit of lattice QCD from the perspective of quantum simulations: For one, as the gauge degrees of freedom can be integrated out exactly in any spatial dimension, Gauss' law is fulfilled implicitly. Secondly, the resulting variables form a discrete set of physical, color-singlet degrees of freedom (mesons and baryons) governed by a local, sparse nearest-neighbor Hamiltonian. Finally, due to the mildness of the sign problem in the Euclidean formulation, cross checks between the formulations can be performed in the extended parameter space of finite $\mu,T$ at \emph{continuous} Euclidean time. It therefore seems natural to map the strong-coupling Hamiltonian onto the degrees of freedom relevant for quantum simulations, starting with the one-flavor, $\beta = 2N_c/g^2 = 0$ case which is the subject of our current study.
%\begin{figure*}[th]
%\centering
%\includegraphics[scale=0.7]{figures/z2_config.pdf}
%\caption{\label{fig:initial_state} }
%\end{figure*}

\section{Background}

The Hamiltonian we study is derived from the continuous Euclidean time limit of strong-coupling lattice QCD with staggered fermions. Strong-coupling lattice QCD can be formulated in dual variables in order to tackle the sign problem, with the full $\mu_B-T$ phase diagram accessible through Monte Carlo simulations. However, because  $T=(a(\beta)N_\tau)^{-1}$ on an isotropic lattice, it requires anisotropic lattices to continuously vary the temperature at fixed inverse gauge coupling $\beta=0$.
The non-perturbative relation between the bare anisotropy $\gamma$, introduced as a prefactor to favor temporal over spatial fermion hops, and the physical anisotropy $\xi\equiv \frac{a}{\at}$, has been determined non-perturbatively for various $\Nt$ and towards the continuous time limit $\Nt\rightarrow\infty$ \cite{Forcrand2018}. The resulting temperature in lattice units is 
$aT=\frac{\xi(\gamma)}{\Nt}$, which simplifies in the continuous time limit to $aT=\kappa\frac{\gamma^2}{\Nt}$ with $\kappa=0.7971(3)$. This is close to the mean-field result where $\kappa=1$.

The first step in the derivation of the Hamiltonian is to expand the partition function in $\gamma^{-1}$, as due to the Grassmann constraint, every spatial meson hop takes away a factor $\gamma^2$ from the temporal contributions \cite{Klegrewe2020,Klegrewe2020a}: 
\begin{widetext}
\begin{align}
\calZ(\gamma, \at \mu_B,\Nt) 
= \calN(\gamma)\sum_{\{k,\ell\}} &\prod_{x\in \LatM}\left(\GC \frac{(3-k_0(x))!}{k_0(x)!}
\prod_{i=1}^{d}\frac{(3-k_i(x))!}{k_i(x)!}\gamma^{-2k_i(x)}
\right)\nn
&\times \prod_{\ell\subset \LatB}\left( \sigma(\ell) \prod_{(x,\mu)\in \ell} 
\Exp{(\delta_{\hmu,+\hat{0}}(\ell)-\delta_{\hmu,-\hat{0}}(\ell))\,\at\mu_B}
\prod_{i=1}^d\lr{\gamma^{-3 \delta_{\mu i}(\ell)}}\right) 
\label{eq:discretePF}
\end{align}
\end{widetext}
with the $(3+1)$-dim.~lattice volume $\Lat$ decomposed into the disjoint union of mesonic sites $\LatM$
and baryonic sites $\LatB$. The integers $k_\mu\in \{0,\ldots 3\}$ are dimers, the selfavoiding loops $\ell$ denote baryons worldlines.
Note that the sign $\sigma(\ell)\in\{+1,-1\}$ can be negative for nontrivial baryon loops, but $\sigma(\ell)=1$ for static baryons. The normalization $\calN(\gamma)$ can be disregarded.
The number of spatial meson hops depends on the temperature, but remains finite at fixed $aT$ as $\Nt\rightarrow \infty$. This results in the suppression of spatial meson hops of orders higher than $\gamma^{-2}$ in units of $a_t$, simplifying \eref{eq:discretePF} significantly:
\begin{widetext}
\begin{align}
\calZ(\gamma, \at\mu_B,\Nt)
%&=  \sum_{
%\substack{
%\left.\{k\}\right|_{\LatM}\\
%\left.\{\omega\}\right|_{\LatSpatB}
%}} \left\{\left(\prod_{x\in \LatM}\GC
%\frac{(3-k_0(x))!}{k_0(x)!}
%\left(\delta_{k_i(x),0}
%+\delta_{k_i(x),1} \frac{1}{\Nc}\gamma^{-2}
%\right)
%\right)\left(\prod_{\vec{x}\in \LatSpatB}  e^{\omega(\vec{x})3 \at \mu_q \Nt}\right)\right\}\nn
&= \sumLatSpat  
\left(
\sum^{\rm GC}_{
\left.\left\{
\substack{k_0\in\{0,\ldots 3\}\\k_i\in\{0,1\}}
\right\}\right|_{\LatM}}
\prod_{(\vec{x},\tau)\in \{ x|k_i(x)=1\}  }\frac{
v(k_0^{-}|k_0^{+})_{(\vec{x},\tau)}
v(k_0^{-}|k_0^{+})_{(\vec{x}+\hat{i},\tau)}
}{\gamma^{2}}
\right)\left( 2 \cosh(\mu_B/T)\right)^{|\LatSpatB|},
\label{eq:ParFuncLargeNt}
\end{align}
\end{widetext}
with $\LatSpatM$ and $\LatSpatB$ denoting the spatial mesonic and baryonic sublattices respectively, which no longer depend on $\tau \in \{0,\ldots \Nt-1\}$. The mesonic contribution has non-trivial weights whenever, at some time index $\tau$, a meson hop between nearest neighbors 
$\langle \vec{x},\vec{x}+\hat{i} \rangle$ occurs. Here 
$v_{(\vec{x},\tau)}$ depends on the the temporal dimers $k_0^{\pm}$ just before/after the spatial dimer location, and due to $k_0^{-}+k_0^{+}+1=3$, the only possible transitions are $v_L\equiv v(0|2)=v(2|0)=1$, $v_T\equiv v(1|1)=2/\sqrt{3}$. In between spatial meson exchange, the temporal dimers at a given site $\vec{x}$ form states that can be identified as meson occupations, $\meson=\{0,\pi,2\pi,3\pi\}$, which correspond to a conserved pion current. 
For the baryonic contribution, spatial hops are suppressed and we have used $\mu_B/T= \at\mu_B \Nt$.

One finally takes the limit $\Nt\rightarrow \infty$ of \eref{eq:ParFuncLargeNt}, by substituting $\gamma^2$ with $aT\Nt$, such that the pion exchange between nearest neighbors can be written as an exponential, and the vertices can be expressed with matrices $\hat{J}^{\pm}$ which raise or lower the meson occupation number, resulting in the following partition function:
\begin{align}
\calZ_{\rm CT}(aT,a\mu_B)&=\Tr_\Hil\left[e^{(-\hat{\calH}+\hat{\calN}a\mu_B)/aT}\right]
\label{eq:CTParFunc}
\end{align}
with the Hamilton and number operators
\begin{align}
\hat{\calH}&=-\frac{1}{2}\sum_{
\langle\vec{x},\vec{y}\rangle}
\left(
\hat{J}^{+}_{\vec{x}}\hat{J}^{-}_{\vec{y}}+
\hat{J}^{-}_{\vec{x}} \hat{J}^{+}_{\vec{y}}
\right),&\hat{\calN}&=\sum_{\vec{x}}\hat{\omega}_x.
\label{eq:hamiltonian}
\end{align}
The corresponding ladder operators $\hat{J}^{-}_x=(\hat{J}^{+}_x)^T$ are:
\begin{align}
\hat{J}^+_x&=\left(
\begin{array}{cccc|cc}
0   & 0   & 0   & 0 &  & \\
v_L & 0   & 0   & 0 &  & \\
0   & v_T & 0   & 0 &  & \\
0   & 0   & v_L & 0 &  & \\
\hline
 &  &  &  & 0 & 0\\
 &  &  &  & 0 & 0\\
\end{array}
\right),\,
\hat{\omega}&=\left(
\begin{array}{cccc|cc}
0 & 0 & 0 & 0 &  & \\
0 & 0 & 0 & 0 &  & \\
0 & 0 & 0 & 0 &  & \\
0 & 0 & 0 & 0 &  & \\
\hline
 &  &  &  & 1 & 0\\
 &  &  &  & 0 & -1\\
\end{array}
\right).
\end{align}
The trace in \eref{eq:CTParFunc} is over the Hilbert space
\begin{align}
\Hil&=\bigotimes_{\vec{x}\in\Sigma} |\hadron_x\rangle ,& |\hadron\rangle&=| \meson,\baryon\rangle =\left(
\begin{array}{c}
0 \\
\pi \\
2\pi \\
3\pi \\
\hline
B^+ \\
B^- \\
\end{array}
\right),
\label{eq:hadron_state}
\end{align}
with $|\hadron\rangle $ denoting the 6 hadronic states per spatial lattice site.
By redefining the meson occupation numbers as $|\spin\rangle=|\meson-3/2\rangle$, a  particle-hole symmetry becomes evident, as the mesonic creation and annihilation operators fulfill the following algebra:
\begin{align}
[\hat{J}^+,\hat{J}^-]&
%=\left(
%\begin{array}{cccc}
%-\hat{v}_\lsteel^2 & 0 & 0 & 0  \\
%0 & \hat{v}_\lsteel^2-\hat{v}_\tsteel^2 & 0 & 0  \\
%0 & 0 & \hat{v}_\tsteel^2-\hat{v}_\lsteel^2 & 0 \\
%0 & 0 & 0 & \hat{v}_\lsteel^2  \\
%\end{array}\right)
=
\left(
\begin{array}{cccc}
-1 & 0 & 0 & 0  \\
0 & -1/3 & 0 & 0  \\
0 & 0 & 1/3 & 0 \\
0 & 0 & 0 & 1  \\
\end{array}\right)\nn
 \hat{J}_1&= \frac{\sqrt{3}}{2}\left(\hat{J}^+ + \hat{J}^-\right),\quad 
 \hat{J}_2= \frac{\sqrt{3}}{2i}\left(\hat{J}^+ - \hat{J}^-\right),\nn
 \hat{J}_3&=i[\hat{J}_1,\hat{J}_2]=\frac{3}{2}[\hat{J}^+,\hat{J}^-], \quad
\hat{J}^2=\frac{15}{4}\Id
%\frac{1}{4}{\rm diag}(2+\Nc^2,2(\Nc-k)(1+k)/\Nc+(\Nc-2)^2,\ldots 2+\Nc^2),
\label{eq:Jis}
\end{align}
This corresponds to a 4-dim.~representation of $\SU(2)$ with $|j,m\rangle\equiv |3/2,\spin \rangle$.\\

This construction can be extended to a $\Nf=2$ Hamiltonian~\cite{Pattanaik2022, Pattanaik2021}, resulting in:
\begin{align}
Z_{\rm CT}(\bareT,\bareMu,\bareI)&=
\Tr_\Hil\left[e^{(-\hat{\calH}+\hat{\calN_B}\bareMu+\hat{\calN_I}\bareI)/\bareT}\right]
\end{align}
where the Hamiltonian is now composed of a sum of 4 contributions, one for each pion 
$\pi\in\{\pi^+,\pi^-,\pi_u,\pi_d\}$:
\begin{align}
  \hat{\calH}&=-\frac{1}{2}\sum_{\langle\vec{x},\vec{y}\rangle}
 \sum_{Q_i \in \{ \pi^+, \pi^-, \pi_U, \pi_D \}} \lr{
 {\hat{J}_{Q_i,\vec{x}}}^+ {\hat{J}_{Q_i,\vec{y}}}^- + {\hat{J}_{Q_i,\vec{x}}}^- {\hat{J}_{Q_i,\vec{y}}}^+% 
 }
 \end{align}
In addition to the baryon number operator $\hat{\calN_B}$, the $\Nf=2$ partition function also contains an isospin number operator $\hat{\calN_I}$.
The number of distinct hadronic states $|\hadron\rangle$ is now 92, of which 50 states are purely mesonic. It turns out that all the matrix elements in $\hat{J}_{Q_i}^{\pm}$ are positive, allowing to study both non-zero baryon and isospin chemical potential.

Another possible extension of the Hamiltonian \eref{eq:hamiltonian} is to perform the strong coupling expansion \cite{Gagliardi2019} before taking the continuous time limit, which results in only temporal plaquettes contributing. Both the $\Nf=2$ formulation and the  inclusion of the leading-order gauge corrections to the Hamiltonian will be discussed in a forthcoming publication.

\begin{figure*}[th]
\scalebox{0.9}{
\Qcircuit @C=1.0em @R=0.2em @!R { \\
	 	\nghost{{q}_{0} :  } & \lstick{{q}_{0} :  } & \qw & \gate{\mathrm{S}} & \gate{\mathrm{H}} & \multigate{3}{\mathrm{diag\_famI}}& \gate{\mathrm{H}} & \gate{\mathrm{S^\dagger}} & \qw & \qw & \qw\\
	 	\nghost{{q}_{1} :  } & \lstick{{q}_{1} :  } & \targ & \qw & \qw & \ghost{\mathrm{diag\_famI}} & \qw & \qw & \targ & \qw & \qw\\
	 	\nghost{{q}_{2} :  } & \lstick{{q}_{2} :  } & \qw & \gate{\mathrm{S}} & \gate{\mathrm{H}} & \ghost{\mathrm{diag\_famI}} & \gate{\mathrm{H}} & \gate{\mathrm{S^\dagger}} & \qw & \qw & \qw\\
	 	\nghost{{q}_{3} :  } & \lstick{{q}_{3} :  } & \ctrl{-2} & \gate{\mathrm{S}} & \gate{\mathrm{H}} & \ghost{\mathrm{diag\_famI}}& \gate{\mathrm{H}} & \gate{\mathrm{S^\dagger}} & \ctrl{-2} & \qw & \qw\\
\\ }}

\scalebox{0.9}{
\Qcircuit @C=1.0em @R=0.2em @!R { \\
	 	\nghost{{q}_{0} :  } & \lstick{{q}_{0} :  } & \qw & \qw & \targ & \gate{\mathrm{R_Z}\,(\mathrm{-\frac{\delta t}{2\sqrt{3}}})} & \qw & \qw & \targ & \gate{\mathrm{R_Z}\,(\mathrm{-\frac{\delta t}{2\sqrt{3}}})} & \targ & \targ & \qw & \qw & \qw\\
	 	\nghost{{q}_{1} :  } & \lstick{{q}_{1} :  } & \qw & \qw & \qw & \targ & \gate{\mathrm{R_Z}\,(\mathrm{-\frac{\delta t}{2\sqrt{3}}})} & \targ & \ctrl{-1} & \qw & \ctrl{-1} & \qw & \qw & \qw & \qw\\
	 	\nghost{{q}_{2} :  } & \lstick{{q}_{2} :  } & \targ & \gate{\mathrm{R_Z}\,(\mathrm{-\frac{\delta t}{2\sqrt{3}}})} & \qw & \ctrl{-1} & \qw & \ctrl{-1} & \qw & \qw & \qw & \qw & \targ & \qw & \qw\\
	 	\nghost{{q}_{3} :  } & \lstick{{q}_{3} :  } & \ctrl{-1} & \qw & \ctrl{-3} & \qw & \qw & \qw & \qw & \qw & \qw & \ctrl{-3} & \ctrl{-1} & \qw & \qw\\
\\ }}
\caption{\label{fig:gate_fam_i} (\emph{Top}) Four qubit quantum gate corresponding to the $\mathrm{Fam}_1$ unitary in the Mesonic Trotter step. (\emph{Bottom}) The diagonal gate {\tt diag\_famI} of the above unitary, decomposed into elementary operations.}
\end{figure*}
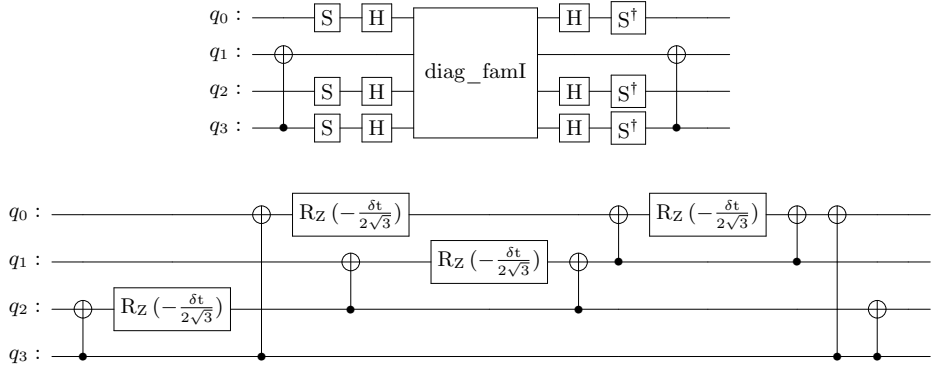

\section{Implementation and Results}

The Hamiltonian $\hat{\calH}$ and the number operator $\hat{\calN}$, as defined in \eref{eq:hamiltonian}, have a block-diagonal structure, with the states $\ket{\hadron}_x$ of the local Hilbert space being a direct \textit{sum} of mesonic and baryonic occupational states, $\ket{\hadron}_x = \ket{\meson}_x \oplus \ket{\baryon}_x$. As a practical consequence, this allows us to qubitize the Hamiltonian \eref{eq:hamiltonian} in the mesonic sector and baryonic sector independently. We begin with the former.
\subsection{Mesonic Gate Sets}
\label{subsect:bosonic_gates}
The four-dimensional local meson state $\ket{\meson}_x =(0, \pi, 2\pi, 3\pi)^T$ can be encoded with two qubits. The Hamiltonian $\hat{\calH}$ represents a nearest-neighbor meson hopping term, seen through the action of the raising and lowering operators, $\hat{J}^+$ and $\hat{J}^-$, respectively, on the local Hilbert space. Such a local, low-dimensional Hamiltonian can be qubitized with standard methods~\cite{Murairi2022} as we summarize below for the case of $1+1$-dim. The method, of course, is completely general and can be applied to higher dimensions.

We start by partitioning the lattice Hamiltonian into mutually commuting even and odd parts
\beq
	\calH = \sum_{x_e} \calH_{x_e} + \sum_{x_o} \calH_{x_o}
\eeq
where $\calH_{x_{e/o}}$ both have the nearest-neighbor $nn$-form of \eref{eq:hamiltonian}, and the labels ``even'' and ``odd'' refer to the bonds of the one-dimensional lattice.
Using the Pauli decomposition $\calH_{x_{e/o}} = \sum_i c_i P_i$ into Pauli strings $P_i \in \{I,X,Y,Z\}^{\otimes 4}$, the remaining tasks are that of partitioning the $P_i$ into sets (``families'') of mutually commuting terms, which can then be simultaneously diagonalized. The Pauli decomposition yields eighteen four-qubit terms (two qubits for the neighboring sites $x$ and $y$, respectively) which can easily be grouped into four families.\footnote{We quote here the result of IBM Qiskit's~\cite{Qiskit2023} {\tt group\_commuting} which uses a graph coloring approach.} The partition of the operators into four families $\mathrm{Fam}_k$ takes the form 
\beq
\mathrm{Fam}_1 &=& \{(IX)_x(YY)_y, (IY)_x(YX)_y + x\leftrightarrow y\; \text{terms} \},\nonumber\\
\mathrm{Fam}_{2} &=& \{XXYY, XYYX, YYXX, YXXY\},\nonumber\\
\mathrm{Fam}_{3} &=& \{IXXX, IYXY, XXIX, XYIY\},\nonumber\\
\mathrm{Fam}_{4} &=& \{YYYY, YXYX, IXIX, IYIY, XXXX, XYXY\}\nonumber ,\\
\eeq
where the big endian convention is used.
The individual terms are multiplied by the coefficients $c_j^{(k)}$ which are listed in the Appendix (\eref{eq:abeliangroup_coeffs}). Each of the $\mathrm{Fam}_{k}$ can be diagonalized simultaneously using the tableau formalism described in~\cite{Berg2020}, yielding the sets
\beq
\mathrm{Diag_{Fam_{1/2}}} &=& \{IIZZ,IZZZ,ZIIZ,ZZIZ\},\nonumber\\
\mathrm{Diag_{Fam_3}} &=& \{IZZI,IZZZ,ZZII,ZZIZ\},\nonumber\\
\mathrm{Diag_{Fam_4}} &=& \{ZZZZ,ZIZZ,IIIZ,IZIZ,IIZZ,IZZZ\}.\nonumber\\
\eeq
The transformations $V_k$ which diagonalize the families of operators
\beq
\exp{\left(i\sum_j c_j^{(k)} P_j^{(k)}\right)} = V_k\exp{\left(i\sum_j \tilde{c}_j^{(k)} \tilde{P}_j^{(k)}\right)}V_k^\dagger
\eeq
are given below in gate form. The remaining step in the derivation of the primitive gate sets for the $\calH_{e/o}$, is the optimization of the exponentiated sum of diagonal terms $\exp{\left(i\sum_j \tilde{c}_j^{(k)} \tilde{P}_j^{(k)}\right)}$. Here, optimization refers to the number of $\mathrm{CNOT}$ gates. Making use of the identity $Z_a\otimes Z_b = \mathrm{CNOT}_{a,b} I_a\otimes Z_b \mathrm{CNOT}_{a,b}$ and taking into account mutual cancellations between neighboring $\mathrm{CNOT}$s, we arrive at the gate set for each of the four exponentiated families. As an example, the relevant gates for the exponentiation of the first family are displayed in Fig.~\ref{fig:gate_fam_i} where the remaining sets are given in the Appendix, Fig.~\ref{fig:gate_fam_app}.

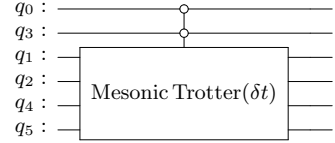
\begin{figure}
\scalebox{0.9}{
\Qcircuit @C=1.0em @R=0.2em @!R { \\
	 	\nghost{{q}_{0} :  } & \lstick{{q}_{0} :  } & \ctrlo{1} & \qw & \qw\\
	 	\nghost{{q}_{3} :  } & \lstick{{q}_{3} :  } & \ctrlo{1} & \qw & \qw\\
	 	\nghost{{q}_{1} :  } & \lstick{{q}_{1} :  } & \multigate{3}{\mathrm{Mesonic\,Trotter}(\delta t)} & \qw & \qw\\
	 	\nghost{{q}_{2} :  } & \lstick{{q}_{2} :  } & \ghost{\mathrm{Mesonic\,Trotter}(\delta t)} & \qw & \qw\\
	 	\nghost{{q}_{4} :  } & \lstick{{q}_{4} :  } & \ghost{\mathrm{Mesonic\,Trotter}(\delta t)}& \qw & \qw\\
	 	\nghost{{q}_{5} :  } & \lstick{{q}_{5} :  } & \ghost{\mathrm{Mesonic\,Trotter}(\delta t)}& \qw & \qw\\
\\ }}
\caption{\label{fig:mesonic_trotter} An example of the two-qubit controlled version of the mesonic Trotter gate where the qubits $(q_0q_1q_2)$ and $(q_3q_4q_5)$ have been arranged for readability.}
\end{figure}

\subsection{Inclusion of Baryons}
Extending our theory from gauge group $\U(3)$ to $\SU(3)$ requires the extension of $|\hadron\rangle=| \meson\rangle$ to $|\hadron\rangle=| \meson, \baryon\rangle$, as defined in \eref{eq:hadron_state}. It is important to note that owing to the limit of infinite gauge coupling \emph{and} continuous time, baryons remain static in our toy theory. Given an initial state $\ket{\psi_0} = \otimes_x \ket{\hadron}_x$, the \emph{time evolution} of bosonic and baryonic sites decouples as $[\calH, \calN] = 0$, with baryons evolving trivially, only leaving an imprint on the surrounding ``pion bath'' through a fixed excluded volume. From the practical side of quantum simulations this implies that the gate set identified so far is sufficient to capture the dynamics of the theory with gauge group $\SU(3)$ in the zero temperature limit.

At nonzero chemical potential or temperature, $\mu_B, T >0$ we will need to extend our register to three qubits per site $x$ to represent all six classical states of $\ket{\hadron}_x = (0, \pi, 2\pi, 3\pi, B^+, B^-)_x^T$, leaving two states as redundancy. As $\calH_{x_{e/o}}$ and $\calN_x$ act on this Hilbert space through a direct sum, we can use the three qubits in the register $(q_0 q_1 q_2)$ and encode the hadronic sector (meson or baryon) of ${\hadron}_x$ in, say, $q_{0}$, thus having it act as \emph{control} bit. A \emph{mesonic} Trotter step thus consists of the gates derived in the last section, Fig.~\ref{fig:gate_fam_i} and Fig.~\ref{fig:gate_fam_app}, but in their controlled version (e.g. on ``0''), with one control bit per site. The unitary $\exp{\left(-i\delta t\calH_{x_{e/o}}\right)}$ acting on pairwise sets of qubits encoding $\ket{\hadron}_x$ and $\ket{\hadron}_y$ of $nn$-pairs $\langle x,y\rangle$, respectively, is hence represented by the symbolic circuit Fig.~\ref{fig:mesonic_trotter}.

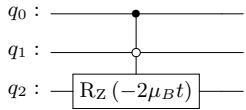
\begin{figure}
\scalebox{0.9}{
\Qcircuit @C=1.0em @R=0.2em @!R { \\
	 	\nghost{{q}_{0} :  } & \lstick{{q}_{0} :  } & \ctrl{1} & \qw & \qw\\
	 	\nghost{{q}_{1} :  } & \lstick{{q}_{1} :  } & \ctrlo{1} & \qw & \qw\\
	 	\nghost{{q}_{2} :  } & \lstick{{q}_{2} :  } & \gate{\mathrm{R_Z}\,(-2\mu_B t)} & \qw & \qw\\
\\ }
}
\caption{\label{fig:baryonic_trotter} Baryonic evolution gate for a single site. The local control bits $q_0$ and $q_1$ encode the hadronic sector and redundant state space, respectively.}
\end{figure}

The baryon evolution happens in diagonal form per site $x$ (i.e. $Z$-term in the Hamiltonian) through the term $\exp{\left(-it\sum_{x}\hat{\omega}_x\right)}$ and hence does not need to be trotterized. We thus only have to take into account local control by $q_0$ (e.g. on ``1''), which encodes the hadronic sector. Locally, to restrict the action of $\exp{\left(-it\omega_x\right)}$ to the $(B^+,B^-)^T$ part of $\ket{\hadron}_x$, we can add a control by $q_1$, thus excluding action on the redundant states of the $2^3$ possible classical states. Fig.~\ref{fig:baryonic_trotter} thus shows the only baryonic gate used in the time evolution of the full ($\SU(3)$) theory.

\begin{figure*}[th]
\includegraphics[scale=.3]{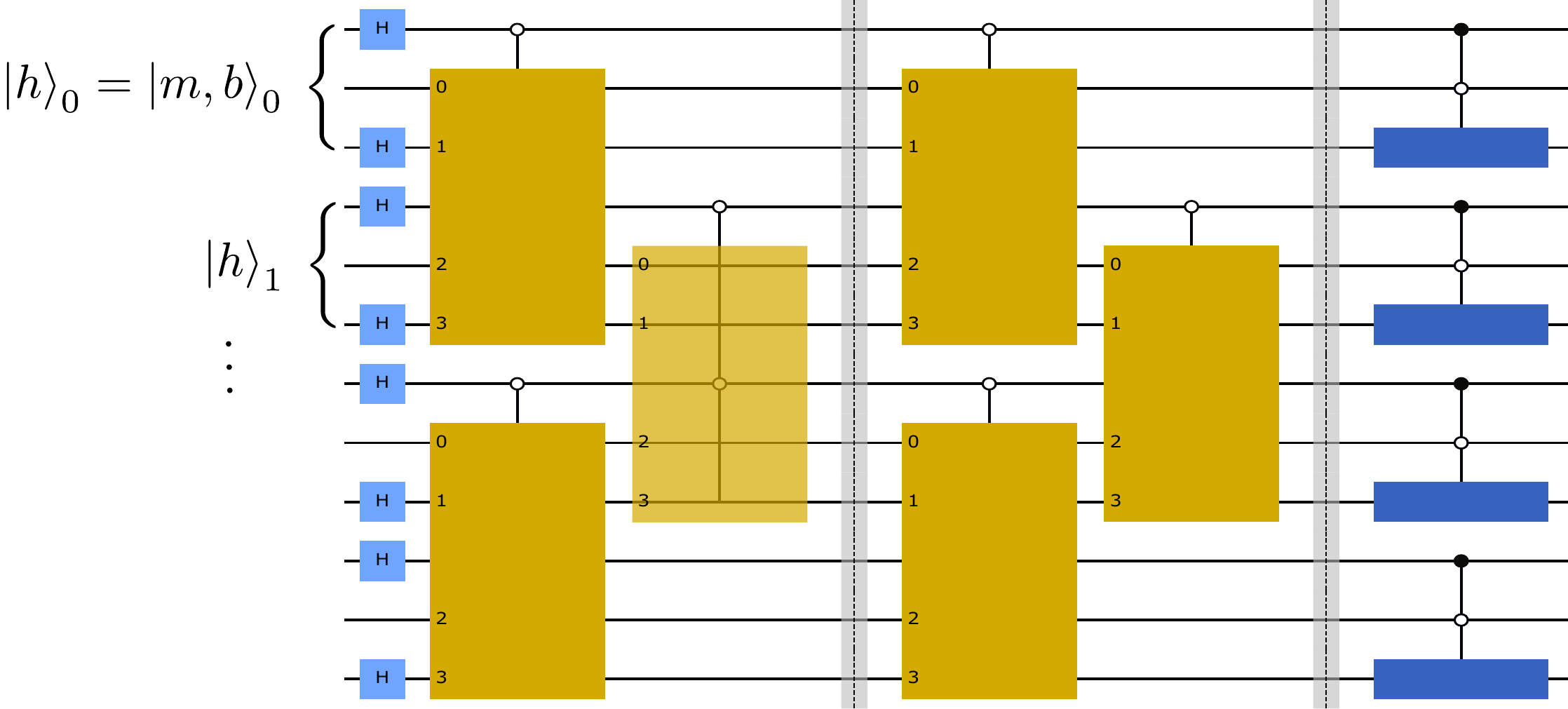}
\caption{\label{fig:trotter_drawing} A symbolic circuit for two mesonic Trotter steps (yellow), followed by a baryonic evolution (blue) for the state $\ket{\psi_0} = (\ket{+}\ket{0}\ket{+})^{\otimes L}$ with linear extent $L = 4$ and open boundary conditions. Each of the mesonic Trotter steps corresponds to a sequential application of the gates defined in Fig.~\ref{fig:gate_fam_i} and Fig.~\ref{fig:gate_fam_app}, where the two-qubit control is visible in the first layer.}
\end{figure*}

\subsection{Trotterized Time Evolution in $d=1+1$}
Trotterizing the time evolution according to
\beq
U(t) = \left(e^{-i\delta t\sum_{x_e} \calH_{x_e}} e^{-i\delta t\sum_{x_o} \calH_{x_o}}\right)^N e^{+it\hat{\calN}\bareMu},
\eeq
where $\, t = \delta t N$, Fig.~\ref{fig:trotter_drawing} depicts the symbolic circuit corresponding to the staggered layers of mesonic Trotter gates, followed by a baryonic evolution for a lattice of dimension $d=1+1$ with linear extent $L=4$, using open boundary conditions for displaying purposes. The initial state chosen, $\ket{\psi_0} = (\ket{+}\ket{0}\ket{+})^{\otimes L}$, corresponds locally to $\ket{\hadron}_x =\frac{1}{2}(1,1,0,0,1,1)^T$, i.e. a superposition of mesonic and baryonic occupation. 

\begin{figure*}[th]
\includegraphics[scale=.53]{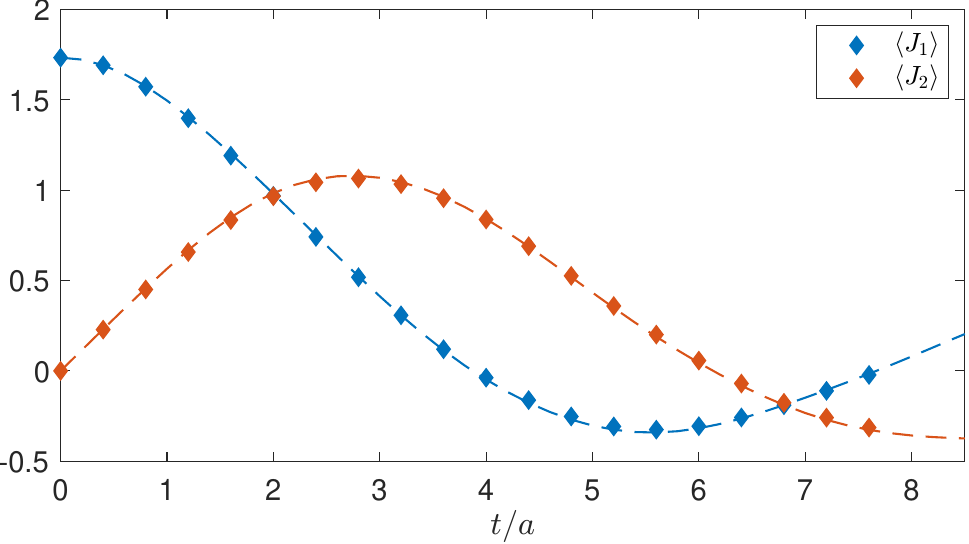} 
\includegraphics[scale=.53]{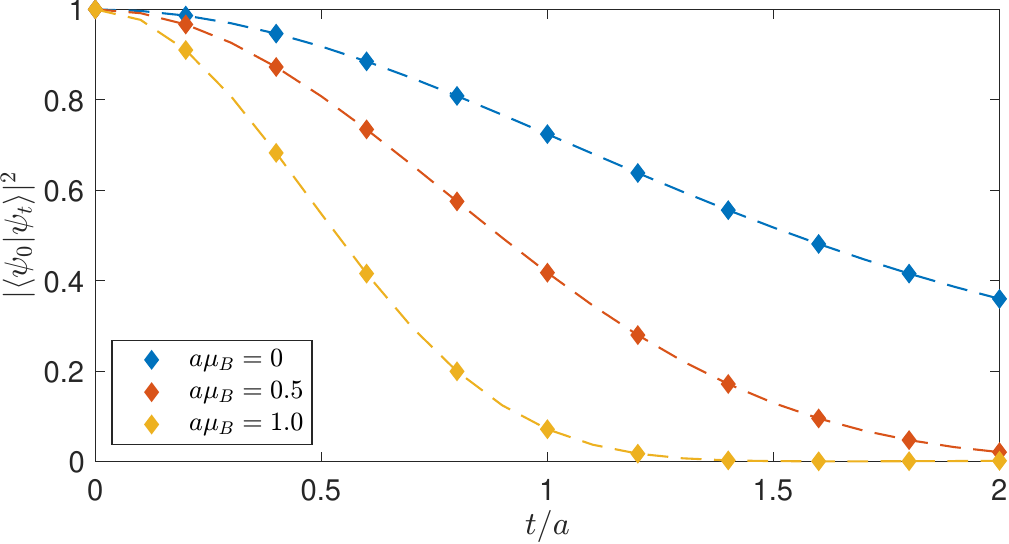}
\caption{\label{fig:Trotter}(\emph{Left}) Comparison of the classically simulated Trotterized expectation value of the mesonic observables  $\hat{J}_{1/2}$ for Trotter step size $\delta t = 0.4$ (diamonds) and the exact results (dashed lines).(\emph{Right}) The wave function overlap $|\braket{\psi_0|\psi(t)}|^2$ for several values of the baryon chemical potential $a\mu_B$ for Trotter step size $\delta t = 0.2$. We used a system of linear extent $L = 4$ and periodic boundary conditions for the initial state $\ket{\psi_0} = (\ket{+}\ket{0}\ket{+})^{\otimes L}$.}
\end{figure*}

To exemplify the correctness of our gate decomposition summarized in Fig.~\ref{fig:trotter_drawing}, we show in Fig.~\ref{fig:Trotter} a comparison of exact and trotterized observables as a function of Trotter time $t$, with initial state $\ket{\psi_0}$ on a lattice of $L = 4$ sites with p.b.c., using $n = 3L$ qubits. Displayed are the results of a noise-free classical simulation, assuming infinite shot statistics, resulting in the time-evolution of expectation values of the mesonic observables $\hat{J}_{1/2}$ (left), defined in \eref{eq:Jis} and the overlap $|\braket{\psi_0|\psi(t)}|^2$ (right) for several values of the baryon chemical potential $a\mu_B$. We observe good agreement with the exact results.

\section{Conclusion and Outlook}

In summary, we have mapped the Hamiltonian for strong-coupling lattice QCD with one flavor of staggered quarks to qubit degrees of freedom by writing down the necessary complete gate set to simulate its dynamics on a quantum computer. This formulation serves as a preparatory step to actual studies of the model on quantum hardware, where it is evident that this will not only allow one to study the dynamics of the model but also its thermal and finite density properties, using e.g. thermal pure quantum states~\cite{Powers2021,Davoudi2022} or the  variational methods described in~\cite{Cerezo2020} and~\cite{Verdon2019}.

Repeating this exercise for $N_f>1$ is another avenue to pursue. Considering the state multiplicity diagram given in~\cite{Pattanaik2022}, one sees that the the purely bosonic sector of the theory (i.e.~$\U(3)$) will require the use of six qubits per lattice site. This perhaps suggests that a mapping of the system to an effective theory or the use of a qudit formulation would be advantageous.

The most physically relevant extension of the current work would be the inclusion of gauge corrections to the Hamiltonian formulation. This would allow one to make contact to other effective theories in the strong-coupling limit~\cite{Ciavarella2023}, and in the longer term, approach the continuum limit $a \to 0$.

\acknowledgments
The work is supported by the Deutsche Forschungsgemeinschaft (DFG, German Research Foundation) through the grant CRC-TR 211 ``Strong-interaction matter under extreme conditions''~--~project number 315477589~--~TRR 211 and by the State of Hesse within the Research Cluster ELEMENTS (Project ID 500/10.006).

\bibliography{pub}

\onecolumngrid

\appendix
\section{Full Gate Sets}
The coefficients $c_i$ of the Pauli string decomposition $\calH_{x_{e/o}} = \sum_i c_i P_i$ are given by $c_i = \mathrm{1}{2^4}\mathrm{tr}(P_i \calH_{e/o})$. After partitioning the entire set of Pauli strings into families of mutually commuting operators, we obtain the following result for the coefficients for the various families
\beq
\label{eq:abeliangroup_coeffs}
{c}^{(1)}_i &=&-\frac{1}{4\sqrt{3}}, \nonumber\\
{c}^{(2)}_i &\in& \{-\phi_{2}, \phi_{2}, -\phi_{2}, \phi_{2}\}, \phi_{2} = \frac{1}{12}\nonumber,\\
{c}^{(3)}_i &\in& \{-\phi_{3}, \phi_{3}, -\phi_{3}, \phi_{3}\}, \phi_{3} = \frac{1}{4\sqrt{3}},\\
{c}^{(4)}_i &\in& \{\phi_4^{(1)}, \phi_4^{(1)}, \phi_4^{(2)}, \phi_4^{(2)}, \phi_4^{(1)}, \phi_4^{(1)}\}, ~\phi_4^{(1)} =-\frac{1}{12}, \phi_4^{(2)} = -\frac{1}{4}\,.\nonumber
\eeq
The diagonalization of the various families may change the sign of the coefficients, yielding
\beq
\label{eq:abeliangroup_coeffs_diag}
\tilde{c}^{(1)}_i,\tilde{c}^{(3)}_i &=& -\frac{1}{4\sqrt{3}}\nonumber,\\
\tilde{c}^{(2)}_i &\in& \{\phi_{2}, -\phi_{2}, -\phi_{2}, \phi_{2}\},\\
\tilde{c}^{(4)}_i &\in& \{\phi_4^{(1)}, -\phi_4^{(1)}, \phi_4^{(2)}, -\phi_4^{(2)}, \phi_4^{(1)}, -\phi_4^{(1)}\}\,.\nonumber
\eeq

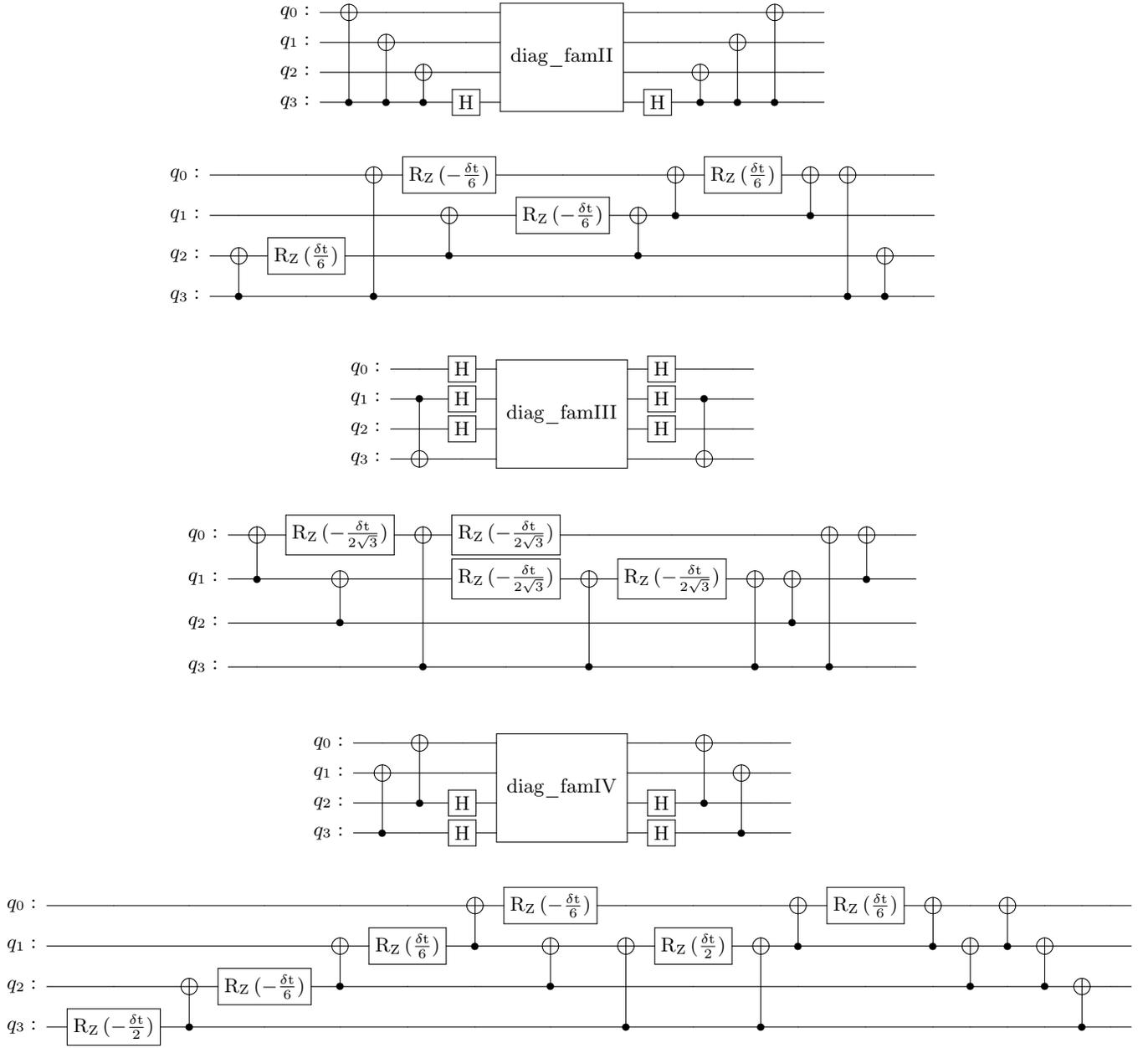
\begin{figure*}[th]
\scalebox{1.0}{
\Qcircuit @C=1.0em @R=0.2em @!R { \\
	 	\nghost{{q}_{0} :  } & \lstick{{q}_{0} :  } & \targ & \qw & \qw & \qw & \multigate{3}{\mathrm{diag\_famII}}& \qw & \qw & \qw & \targ & \qw & \qw\\
	 	\nghost{{q}_{1} :  } & \lstick{{q}_{1} :  } & \qw & \targ & \qw & \qw & \ghost{\mathrm{diag\_famII}} & \qw & \qw & \targ & \qw & \qw & \qw\\
	 	\nghost{{q}_{2} :  } & \lstick{{q}_{2} :  } & \qw & \qw & \targ & \qw & \ghost{\mathrm{diag\_famII}} & \qw & \targ & \qw & \qw & \qw & \qw\\
	 	\nghost{{q}_{3} :  } & \lstick{{q}_{3} :  } & \ctrl{-3} & \ctrl{-2} & \ctrl{-1} & \gate{\mathrm{H}} & \ghost{\mathrm{diag\_famII}} & \gate{\mathrm{H}} & \ctrl{-1} & \ctrl{-2} & \ctrl{-3} & \qw & \qw\\
\\ }}

\scalebox{1.0}{
\Qcircuit @C=1.0em @R=0.2em @!R { \\
	 	\nghost{{q}_{0} :  } & \lstick{{q}_{0} :  } & \qw & \qw & \targ & \gate{\mathrm{R_Z}\,(\mathrm{-\frac{\delta t}{6}})} & \qw & \qw & \targ & \gate{\mathrm{R_Z}\,(\mathrm{\frac{\delta t}{6}})} & \targ & \targ & \qw & \qw & \qw\\
	 	\nghost{{q}_{1} :  } & \lstick{{q}_{1} :  } & \qw & \qw & \qw & \targ & \gate{\mathrm{R_Z}\,(\mathrm{-\frac{\delta t}{6}})} & \targ & \ctrl{-1} & \qw & \ctrl{-1} & \qw & \qw & \qw & \qw\\
	 	\nghost{{q}_{2} :  } & \lstick{{q}_{2} :  } & \targ & \gate{\mathrm{R_Z}\,(\mathrm{\frac{\delta t}{6}})} & \qw & \ctrl{-1} & \qw & \ctrl{-1} & \qw & \qw & \qw & \qw & \targ & \qw & \qw\\
	 	\nghost{{q}_{3} :  } & \lstick{{q}_{3} :  } & \ctrl{-1} & \qw & \ctrl{-3} & \qw & \qw & \qw & \qw & \qw & \qw & \ctrl{-3} & \ctrl{-1} & \qw & \qw\\
\\ }}

\scalebox{1.0}{
\Qcircuit @C=1.0em @R=0.2em @!R { \\
	 	\nghost{{q}_{0} :  } & \lstick{{q}_{0} :  } & \qw & \gate{\mathrm{H}} & \multigate{3}{\mathrm{diag\_famIII}}& \gate{\mathrm{H}} & \qw & \qw & \qw\\
	 	\nghost{{q}_{1} :  } & \lstick{{q}_{1} :  } & \ctrl{2} & \gate{\mathrm{H}} & \ghost{\mathrm{diag\_famIII}} & \gate{\mathrm{H}} & \ctrl{2} & \qw & \qw\\
	 	\nghost{{q}_{2} :  } & \lstick{{q}_{2} :  } & \qw & \gate{\mathrm{H}} & \ghost{\mathrm{diag\_famIII}} & \gate{\mathrm{H}} & \qw & \qw & \qw\\
	 	\nghost{{q}_{3} :  } & \lstick{{q}_{3} :  } & \targ & \qw & \ghost{\mathrm{diag\_famIII}} & \qw & \targ & \qw & \qw\\
\\ }}

\scalebox{1.0}{
\Qcircuit @C=1.0em @R=0.2em @!R { \\
	 	\nghost{{q}_{0} :  } & \lstick{{q}_{0} :  } & \targ & \gate{\mathrm{R_Z}\,(\mathrm{-\frac{\delta t}{2\sqrt{3}}})} & \targ & \gate{\mathrm{R_Z}\,(\mathrm{-\frac{\delta t}{2\sqrt{3}}})} & \qw & \qw & \qw & \qw & \targ & \targ & \qw & \qw\\
	 	\nghost{{q}_{1} :  } & \lstick{{q}_{1} :  } & \ctrl{-1} & \targ & \qw & \gate{\mathrm{R_Z}\,(\mathrm{-\frac{\delta t}{2\sqrt{3}}})} & \targ & \gate{\mathrm{R_Z}\,(\mathrm{-\frac{\delta t}{2\sqrt{3}}})} & \targ & \targ & \qw & \ctrl{-1} & \qw & \qw\\
	 	\nghost{{q}_{2} :  } & \lstick{{q}_{2} :  } & \qw & \ctrl{-1} & \qw & \qw & \qw & \qw & \qw & \ctrl{-1} & \qw & \qw & \qw & \qw\\
	 	\nghost{{q}_{3} :  } & \lstick{{q}_{3} :  } & \qw & \qw & \ctrl{-3} & \qw & \ctrl{-2} & \qw & \ctrl{-2} & \qw & \ctrl{-3} & \qw & \qw & \qw\\
\\ }}

\scalebox{1.0}{
\Qcircuit @C=1.0em @R=0.2em @!R { \\
	 	\nghost{{q}_{0} :  } & \lstick{{q}_{0} :  } & \qw & \targ & \qw & \multigate{3}{\mathrm{diag\_famIV}}& \qw & \targ & \qw & \qw & \qw\\
	 	\nghost{{q}_{1} :  } & \lstick{{q}_{1} :  } & \targ & \qw & \qw & \ghost{\mathrm{diag\_famIV}} & \qw & \qw & \targ & \qw & \qw\\
	 	\nghost{{q}_{2} :  } & \lstick{{q}_{2} :  } & \qw & \ctrl{-2} & \gate{\mathrm{H}} & \ghost{\mathrm{diag\_famIV}} & \gate{\mathrm{H}} & \ctrl{-2} & \qw & \qw & \qw\\
	 	\nghost{{q}_{3} :  } & \lstick{{q}_{3} :  } & \ctrl{-2} & \qw & \gate{\mathrm{H}} & \ghost{\mathrm{diag\_famIV}} & \gate{\mathrm{H}} & \qw & \ctrl{-2} & \qw & \qw\\
\\ }}

\scalebox{1.0}{
\Qcircuit @C=1.0em @R=0.2em @!R { \\
	 	\nghost{{q}_{0} :  } & \lstick{{q}_{0} :  } & \qw & \qw & \qw & \qw & \qw & \targ & \gate{\mathrm{R_Z}\,(\mathrm{-\frac{\delta t}{6}})} & \qw & \qw & \qw & \targ & \gate{\mathrm{R_Z}\,(\mathrm{\frac{\delta t}{6}})} & \targ & \qw & \targ & \qw & \qw & \qw & \qw\\
	 	\nghost{{q}_{1} :  } & \lstick{{q}_{1} :  } & \qw & \qw & \qw & \targ & \gate{\mathrm{R_Z}\,(\mathrm{\frac{\delta t}{6}})} & \ctrl{-1} & \targ & \targ & \gate{\mathrm{R_Z}\,(\mathrm{\frac{\delta t}{2}})} & \targ & \ctrl{-1} & \qw & \ctrl{-1} & \targ & \ctrl{-1} & \targ & \qw & \qw & \qw\\
	 	\nghost{{q}_{2} :  } & \lstick{{q}_{2} :  } & \qw & \targ & \gate{\mathrm{R_Z}\,(\mathrm{-\frac{\delta t}{6}})} & \ctrl{-1} & \qw & \qw & \ctrl{-1} & \qw & \qw & \qw & \qw & \qw & \qw & \ctrl{-1} & \qw & \ctrl{-1} & \targ & \qw & \qw\\
	 	\nghost{{q}_{3} :  } & \lstick{{q}_{3} :  } & \gate{\mathrm{R_Z}\,(\mathrm{-\frac{\delta t}{2}})} & \ctrl{-1} & \qw & \qw & \qw & \qw & \qw & \ctrl{-2} & \qw & \ctrl{-2} & \qw & \qw & \qw & \qw & \qw & \qw & \ctrl{-1} & \qw & \qw\\
\\ }}

\caption{\label{fig:gate_fam_app} The four-qubit quantum gates corresponding to the unitary operators $\mathrm{Fam}_{2/3/4}$ in the mesonic Trotter step along with the respective decomposition of the diagonal gates {\tt diag\_famII/III/IV}.}
\end{figure*}

\end{document}